# Teaching Einsteinian Physics at Schools: Part 1, Models and Analogies for Relativity

**Tejinder Kaur[1], David Blair[1], John Moschilla[1], Warren Stannard[1] and Marjan Zadnik[1]**

[1]The University of Western Australia, 35 Stirling Highway, Crawley, WA 6009, Australia.

E-mail: tkaur868@gmail.com

**Abstract**
The Einstein-First project aims to change the paradigm of school science teaching through the introduction of modern Einsteinian concepts of space and time, gravity and quanta at an early age. These concepts are rarely taught to school students despite their central importance to modern science and technology. The key to implementing the Einstein-First curriculum is the development of appropriate models and analogies. This paper is the first part of a three-paper series. It presents the conceptual foundation of our approach, based on simple physical models and analogies, followed by a detailed description of the models and analogies used to teach concepts of general and special relativity. Two accompanying papers address the teaching of quantum physics (Part 2) and research outcomes (Part 3).

## 1. Introduction

The term 'Einsteinian Physics' refers to the theories of Special and General Relativity developed by Einstein,[1] and Quantum Physics that had its origin in the discovery of photons by Einstein[2] and Planck.[3] These theories are two of the major pillars of modern physics. They give the best description to date of the universe in which we live. Unfortunately, most of the concepts of Einsteinian physics are rarely part of the school science curriculum worldwide.[4],[5],[6] When aspects of Einsteinian physics are introduced, it is usually only for specialist physics students.

It is generally believed that Einsteinian physics is difficult, highly mathematical and only suitable for gifted students.[7] The question of the difficulty of Einsteinian physics can be challenged. Newtonian physics can also be extremely mathematically complex, but those aspects are not included in the school science curriculum. Einsteinian physics can be taught from an observational point of view, without recourse to complex abstract mathematics, either without formulae or with formulae no more complicated than those of Newtonian physics.

Many researchers have developed different approaches for introducing the concepts of special and general relativity and quantum physics.[8],[9],[10],[11] The program described here enables almost all learning to take place through practical activities based on models and analogies.

In this paper, Section 2 discusses the conceptual differences between Newtonian and Einsteinian physics, and the importance of models and analogies in physics education. Section 3 discusses the models and analogies under three subheadings: general relativity, geometry in curved space and special relativity, and how they are used in a systematic introduction to Einsteinian physics. Appendix 1 gives detailed instructions to allow teachers to replicate our models.

## 2. The importance of models and analogies for teaching physics
Models and analogies are particularly important for introducing Einsteinian physics. Previous research has shown that the use of models and analogies in classroom teaching provides a route for students to gain conceptual understanding.[12],[13] These are essential tools not only for a scientific description of the world but also for those things which are not directly perceivable by the human senses.[14] When models and analogies are used, it is important to understand and emphasise the limitations and differences between the analogous system and the real world. In the process of using models and analogies, we develop ways of understanding what the universe is *like*, while we can rarely claim to know what it *is*.

The first trials of the Einstein-First research project began with Year 6 primary school students. This work was based on the hypothesis that the worldview of Einsteinian physics would be much more acceptable to young and malleable minds because it did not contradict their prior knowledge.[15]

Furthermore, if students are to acquire our best understanding of the universe, we propose that it is better that they learn the fundamental concepts of Einsteinian physics before those of Newtonian physics. Once they have grasped the Einsteinian concepts, they can easily learn that most of the time, on Earth, we can treat time as being absolute, space as being flat, and gravity as a force field coming out of the Earth.

To teach Einsteinian physics requires the development of suitable models and analogies. The challenge for the work described here was to create suitable models and learning sequences to allow students to develop a deep understanding of the concepts of Einsteinian physics.

Previous study has shown that working in groups and peer discussion are important in enhancing students' understanding and thinking skills.[16] We used these approaches in the Einstein-First project. The Einstein-First program was trialled in different schools with different age groups. The content was delivered through models and analogies combined with on-screen presentations. The primary component consisted of small group work with the physical model systems which were all designed for activity-based learning. When possible, we asked students to use a video recording of experiments and to plot the results using frame-by-frame analysis. In the next section, we present an analysis of the models we have used.

### 3. Models and analogies for teaching Einsteinian physics
This section introduces the models and analogies that we use to encapsulate various concepts of Einsteinian physics. It provides the relationship between the physics we intend to teach and the models we use for the teaching. In this paper, we focus specifically on the models used for introducing general and special relativity.

### 3.1 General Relativity
An understanding of Einstein's general theory of relativity[17] requires students to absorb the following concepts: a) Mass causes curvature in space-time, and b) Freely falling bodies including photons, follow the shortest paths in space-time.

We suggest that there are two key reasons that general relativity has not entered the mainstream of education. Firstly, the above concepts are difficult to visualise and comprehend without a methodical introduction to the ideas. Secondly, general relativity is normally presented from a mathematical point of view, with mathematics that is far beyond normal high school levels. We avoid the complex mathematics completely by making use of simple formulae that refer to black holes, weakly curved space and time dilation.[18] These formulae are no more complex than the standard formulae used in Newtonian physics. Because we see some value in rote learning, we also encourage students to memorise John Wheeler's aphorism "*Matter tells space-time how to curve; Space-time tells matter how to move.*"[19] These words encapsulate the central concept of general relativity, and we use them to frame a model designed to illustrate this relationship. In this model, we use a membrane made from the sports clothing material called lycra (or spandex) to represent deformable space and objects of various mass can be placed on this membrane to simulate gravitational motion in other objects.

This model, which is sometimes known as 'the space-time simulator', has been used since 2002[20],[21] and was independently developed at the University of Western Australia in 2005.[22] In 2013, White described its use: "A handful of marbles and a large sheet of spandex can make a versatile, convenient, and alluring interactive science apparatus for the classroom".[23] In 2014, Wu et al. tested the model theoretically and experimentally, showing that it accurately represents planetary orbits, but breaks down for more complicated orbital systems, for instance equal mass binary systems.[24] Because of the intrinsic non-Euclidean geometry of the model, it also manifests the geodetic effects of general relativity. It is interesting that even though the model was first developed for demonstrating Newtonian physics, it is actually much better suited for general relativity, except for the important limitations of the model that we discuss later (see Section 3.1.9).

Here we present a systematic approach based on the use of this model that takes students through a learning progression designed to create a deep conceptual understanding of general relativity. We introduce concepts step by step using qualitative and quantitative experiments, depending on the age group. The following concepts are covered: 1) the relationship between matter and (curved) space; 2) "space tells matter how to move", 3) mapping the shape of space using photon trajectories, 4) gravitational lensing, 5) testing Newton's laws of gravitation, as well as 6) Kepler's third law of planetary motion, 7) geodetic precession, and 8) binary star systems and gravitational waves.

*3.1.1 Matter and space: understanding how matter creates curved space*
An elastic membrane stretched across an elevated wooden frame is used to represent space. In its neutral state, it represents flat space as shown in Figure 1(a). Students measure its flatness with a ruler and note that a single ball generally experiences no transverse forces. Students are asked to check how well this works and note that the edges 'repel' the ball (this is a limitation of our model).

Adding masses to this membrane will cause it to distort into the third dimension, thus representing curved two-dimensional space. By adding increasing amounts of mass, students can measure the deformation of the membrane using rulers, and draw a graph of deformation versus number of total mass (or number of balls). We ask students to observe the fact that as space is deformed the area of the space increases. If white dots are marked on the membrane, students also observe that the spacing between dots increases. A plot of this relationship is shown in Figure 1(b).

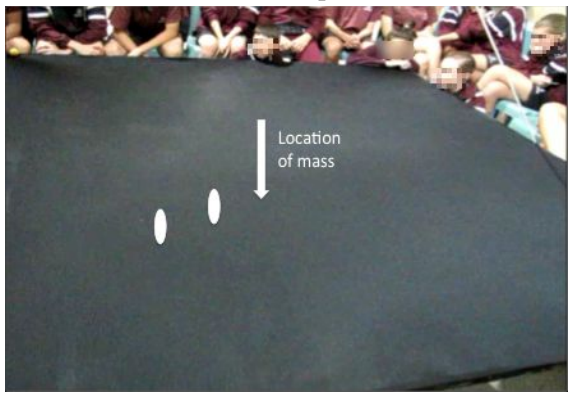
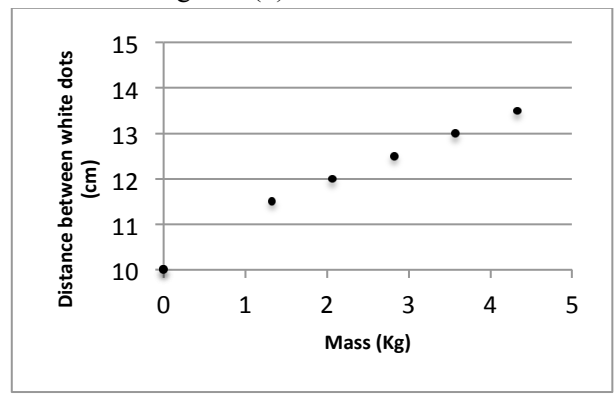

Figure 1(a). Our space-time simulator in its neutral state representing flat two-dimensional space. In the presence of mass, the membrane deforms. For objects on curved space, the force law of gravity is similar to Newton's law of gravitation (see section 3.1.5.)

Figure 1(b). This graph shows typical measurements of the stretching of space between two white dots with different central mass. Initially when there is no central mass, the distance between the dots in Figure 1(a) is 10 cm. With increasing mass (and increasing curvature), the distance between the dots increases.

*3.1.2 Gravitational motion in curved space: "Space tells matter how to move"*

Students can examine how masses (balls) placed on this membrane naturally "attract" each other. First, students observe pairs of balls rolling together and then examine the acceleration of small test balls as a function of the total number of balls that make up the central mass. Students can even observe the universality of free fall (the equivalence principle) by using a large central mass, and then observing the acceleration of pairs of balls of different masses (Figure 2). As long as they are homogeneous solid balls, their in-fall time depends only weakly on their mass. Results can be quantitative or qualitative (ie. graphical and mathematically or explanation-based), depending on the level of the students.

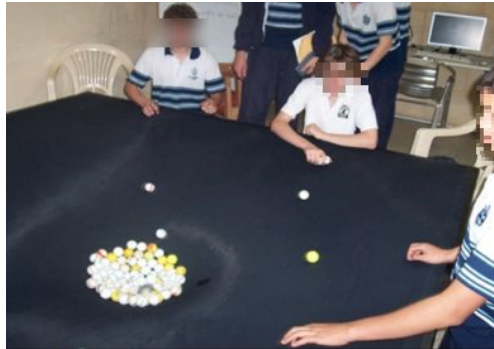

Figure 2. Students orbiting golf balls around a central mass. The balls orbit the central mass due to the curvature of the membrane caused by the central mass. These orbits resemble the orbit of planets around stars. The greater the central mass, the greater the curvature of space-time and the stronger the force of gravitation exerted on the orbiting bodies.

*3.1.3 Mapping the shape of space with photon trajectories*

Having created large deformations, students can now explore the meaning of straight lines in curved space. Students are asked to think of the meaning of two-dimensional curved space. They are asked to imagine themselves as two-dimensional organisms (such as microbes or flatworms) that only experience the two dimensions of the membrane surface. We now introduce photons: we will think of them as projectiles that do not feel any force of gravity, and have no steering. They only know the space they are in (in the two-dimensional case, we mean "on"!). They travel in a straight line just like a toy car without steering always follows a straight line. For these reasons, we adopt a toy car as an analog 'photon' or light beam, such as a beam of starlight approaching us.

In flat space (an empty membrane or a flat floor) students can confirm that the cars follow straight paths. Repeating this on curved space, students will see that the cars follow a curved path. The deformation of the trajectory increases as the deformation of the space increases. Parallel trajectories can be seen to converge and cross at different distances beyond the central mass. The observed deviations define the shape of the curved space around the central massive body.

A very important insight can be obtained by replacing the toy cars with fast moving balls. For slow motion, balls follow orbits which are easily understood as arising from a central force. At high speed, the gravitational deflection is minimised in the same way that it is eliminated using the toy car. Students observe that at high speed the ball has the same deflection as toy car, because it cannot avoid the spatial curvature effects.

The intersection of parallel paths is a violation of Euclid's fifth postulate — also called the parallel postulate — which states that parallel lines never meet. This is observed in astronomical observations which we discuss below.

*3.1.4 Astronomical observations of curved space: Gravitational lensing*

The trajectories that we used above to map the shape of space are an example of gravitational lensing. In general, the bending of light near massive objects such as the Sun, is known as gravitational lensing. It was first observed rather inaccurately by Eddington in 1919[25], then measured more accurately as $1.82 \pm 0.20$ arcsecond[26] by the Wallal eclipse expedition in Western Australia in 1922.[27] Today the value has been measured as $1.66 \pm 0.18$ arcsecond[28]. Also today astronomical photographs easily available on the internet show the dramatic distortions that gravitational lenses create when the light from distant galaxies is distorted by galaxies in the forground as discussed below.

The purpose of this topic is to extend students understanding of deflection in two-dimensions, enable them to understand these astronomical images.

Using pairs of toy cars in parallel trajectories at opposite sides of the central mass (see Figure 3(a)) students can observe focal points that depend on the impact parameter 'b' which is the transverse distance shown in Figure 3(a). Students need to understand the incoming parallel paths as light from a single very distant light source. Then they need to consider themselves as observers looking from behind the lensing object. At each of the points marked with arrows behind the mass, they see the same light from two different directions. The single source has become double. In three-dimensions it can become multiple and sometimes a single background source is distorted into huge arcs. We complete this activity by asking students to search the internet for images of gravitational lensing such as Figure 3(b).

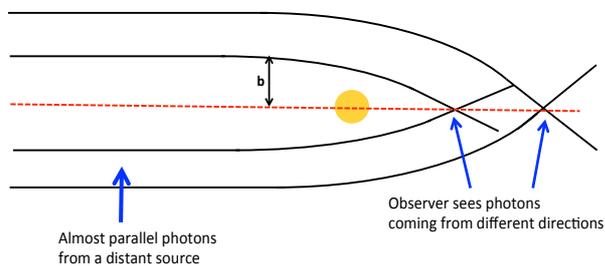

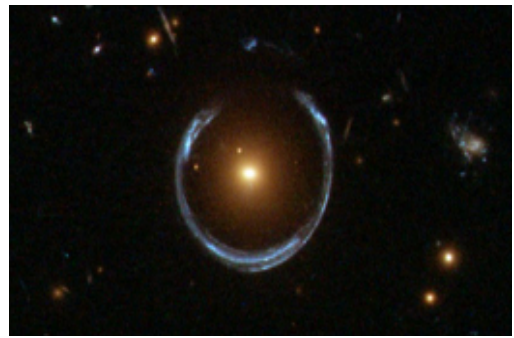

Figure 3(a). Diagram of lensing trajectories which can be observed using toy cars on the two-dimensional membrane. The impact parameter is denoted by 'b'.

Figure 3(b).[1] Gravitational lensing of light from a distant galaxy by a much nearer massive galaxy, creates a nearly complete Einstein ring.

The mathematics of gravitational lensing is simple. The deflection angle $\propto$, of light of light passing an object of mass M is given by

$$\propto = \frac{4GM}{c^2\,b}$$

Where b is the impact parameter, c is the speed of light (c = 3.0 x 10$^8$ ms$^{-1}$), and G is the gravitational constant (G = 6.674 x 10$^{-11}$ m$^3$ kg$^{-1}$ s$^{-2}$). Mathematically competent students can use this formula to estimate the mass of a lensing galaxy assuming an impact parameter of say 10$^5$ light years.

*3.1.5 Investigating Newtonian gravity: deriving a force law for the membrane*

While the spacetime simulator is usually employed to illustrate curved spacetime, it can also be used to investigate the Newtonian conception of gravitation. The activity described below allows students to measure, and plot, Newton's force law of gravitation in the context of space-time curvature.

For this activity, students will need a toy car with fixed steering (with a mass approximately 10% of the central mass). Letting this car roll from the perimeter of the membrane toward the central mass, students will observe the car accelerate. It is useful if students video the motion and plot the spacetime trajectory. The next step of this activity involves measuring the force of 'gravity' exerted on the car as a function of distance from the central mass (see Figure 4(a)). This can be done by connecting the car to a small digital spring balance with a piece of light string. By placing the car at regularly spaced distances from the central mass, students can measure the tensile force exerted by the car as a function of distance from the central mass. (Note that our car has a space for adding mass.)

This relationship can be plotted (see Figure 4(b)) and represents the 'force of gravity' exerted by the membrane at increasing radius from the central mass. From this plot, a force law of gravitation for the membrane can be derived. Just like real Newtonian gravitation, this force law take the form

$$F = \frac{Am_1 m_2}{r^x}$$

where $A$ is a membrane elasticity factor (which in this model corresponds to the gravitational constant G), $m_1$ is the mass of test mass (toy car), $m_2$ is the mass of the central mass, and r is the distance from the central mass on a curved surface.

Using Excel or a similar spreadsheet it is easy to match the data of Figure 4(b) with curves of the form $F = Br^{-0.6}$ where the constant B can be selected by trial and error (formally $B = Am_1 m_2$). Thus the law of gravitation for the membrane is $F = Am_1 m_2 / r^{0.6}$.

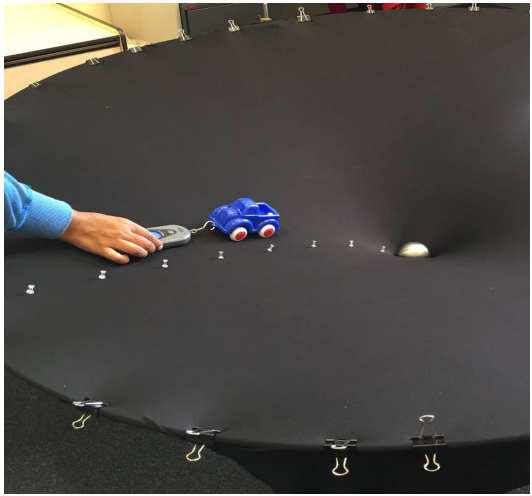

Figure 4(a). A spring balance is used to measure the force of 'gravitation' exerted on a test mass at various distances from the central mass. Since the curvature of the membrane is greatest near the central mass, so is the force exerted on the test mass.

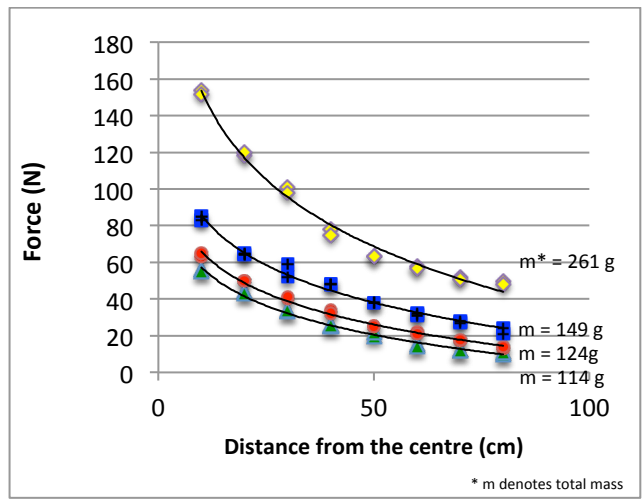

Figure 4(b). Graph of the force of 'gravitation' exerted on test masses of various mass, with increasing distance from the central mass. As expected, the force exerted is greatest for heavier orbiting masses, and when the orbiting mass is closest to the central mass. The fitted curves follow the $r^{-0.6}$ law.

*3.1.6 Measuring orbits: Kepler's laws*

Kepler's third law states that "the square of the orbital period of a planet is proportional to the cube of the semi-major axis of its orbit". Mathematically, $T^2 \propto r^3$, where T is the orbital period and r is the semi-major axis. This law relates the period of a planetary orbit to its distance from the Sun. This law is easily derived in the Newtonian approximation by equating $G\frac{m_1 m_2}{r^2}$ with $\frac{m_1 v^2}{r}$ (see Appendix 2). While this is true to a very good approximation for planetary orbits, it will not be true on the elastic membrane because of the different force law ($r^{-0.6}$ as opposed to $r^{-2}$).

It is instructional for students to appreciate this breakdown in the model by testing Kepler's third law on the membrane. Students start this activity by rolling balls in large circular orbits around the central mass. In reality the balls always spiral towards the central mass, so circular orbits are only approximate. Students either use stop-watches or record videos of the contracting orbits. Using frame-by-frame analysis of the video (and knowing the frame rate of their recording device), they can measure the orbital period for different radii, which can be graphed (see Figure 5). This can be compared against theoretical predictions. Students should observe that the period of the orbit is the same whether the ball is a golf ball, marble or ball bearing.

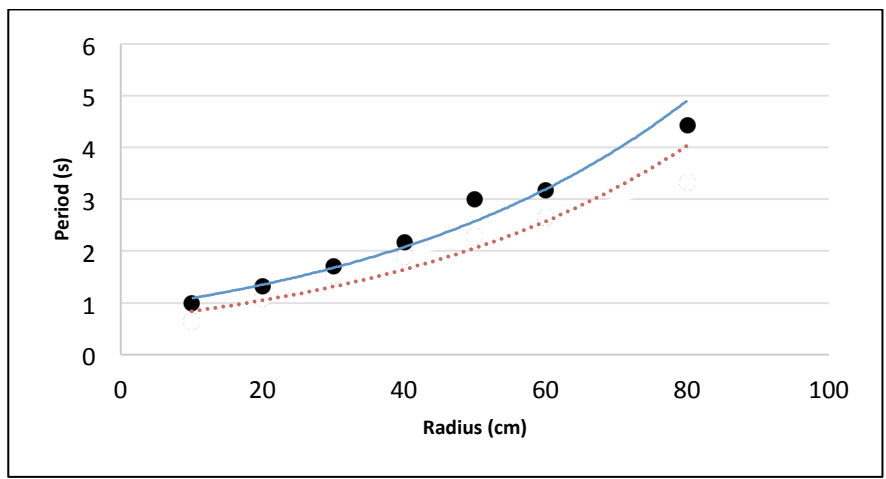

Figure 5. A typical plot of a period versus radius for balls orbiting a central mass on the space-time simulator. A curve has been fitted to the data (solid blue line). The theoretical curve $T \propto Ar^{0.8}$ (derived in Appendix 3) for orbits on the space-time simulator is shown for comparison (dotted orange line).

Two other beautiful classical orbital phenomena can easily be observed on the spacetime simulator. The first is three body interactions, by which a pair of close spaced rolling balls transfer energy between them, causing one orbit to rapidly decay while another is boosted. Secondly, if a cluster of marbles is rolled in an orbit, the system stays gravitational bound until it experiences strong gravity gradient from the central object. Then it is broken up (as observed for comet Shoemaker-Levy [29]) and becomes a long string of objects.

*3.1.7 The effect of curved space on orbits: geodetic precession*

In the previous activity, we used the elastic membrane to investigate Newtonian gravity through Kepler's third law. Now we are going to consider how in curved space Kepler's second law and third law of planetary motion are incompatible. Keplers second law states that planetary orbits sweep out equal areas in equal times as illustrated in Figure 6(a). Using elliptical orbits students observe high velocity near the central mass and low velocity farther from the central mass.

Kepler's second and third law can be easily derived from Newton's Law of Gravitation, in flat space. However, in the curved geometry of the elastic membrane, these two formulations of Keplerian dynamics can be easily shown to be non-commensurate because the area A(r) of an orbit increases differently from its circumference C(r), as you increase the deformation of the membrane.

Specifically, the area to circumference ratio $\frac{A(R)}{C(R)}$ of a circle of radius R is no longer equal to $\frac{R}{2}$ as it is for circles in Euclidean geometry. This means that an elliptical orbital pattern cannot be stationary in curved space, but must precess. This phenomenon is called geodetic precession, and has been observed in numerous orbital systems (for example precession in binary pulsars). It was observed for Mercury during the 19th century, and provided the first evidence that Newton's law of gravitation was not a complete description of gravity.[30] To observe this phenomenon, we ask students to record a video of elliptical orbits, and observe the characteristic daisy-like precession pattern shown in Figure 6(b). Strong precession effects can easily be seen, but generally losses in the membrane mean that only two or three precessing orbits can be observed. A challenge for students is to obtain the best video of precession.

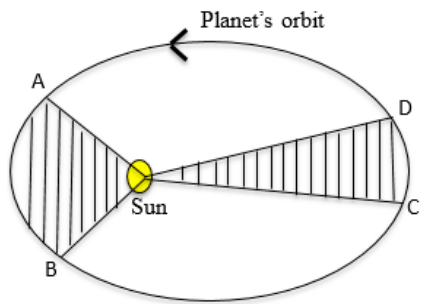

Figure 6(a). According to Kepler's second law, the line between the Sun and the planet sweeps equal areas of space in equal lengths of time. This is observed for masses orbiting a central mass whereby the speed of the orbiting mass visibly increases as it nears the central mass and slows down as it moves further away.

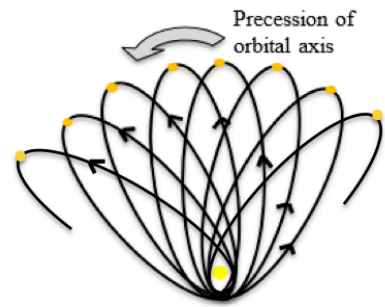

Figure 6(b). A diagram of geodetic precession of the orbit of a planet around a star. This can be observed on the elastic membrane for elliptical orbits of small masses orbiting heavy central mass whereby the farthest point from the central mass shifts position with successive orbits. Due to energy losses on the membrane, only a few precessions can be observed before the orbit deteriorates.

*3.1.8 Binary star systems and gravitational waves*
Ideally, we would like to observe the rippling space-time vibration produced in binary orbits, that appear in many computer simulations such as simulations available on the LIGO website.[31] The motion of binary star systems is the most difficult phenomenon to observe in membrane experiments because the balls experience strong drag from stretching friction. This friction arises from the sliding that occurs when the membrane surface is forced to conform to the spherical shape of the ball.

In spite of the difficulties, it is very instructive for students to experiment with binary systems on the membrane. To observe binary star systems, students need to experiment with pairs of balls of different mass ratios. For a small mass orbiting a large mass, the dynamics is dominated by the large mass. Friction is small. However, the large mass experiences strong stretching friction and fails to show the gravitational recoil that occurs in real gravitating systems. For two equal masses, both have large stretching friction and their mutual orbit collapses rapidly. Measurement of radial motion recoil is used by astronomers to detect planets, but is very difficult to observe on an elastic membrane.

The purpose of this activity is to enable students to understand the common motion about the centre of mass and the recoil of the larger mass. Secondly, students should be able to visualise how wave-like vibrations (i.e. gravitational waves) could be created in the membrane. The difficulty of creating isolated binary orbits on an elastic membrane becomes an interesting challenge. It can be combined with video simulations to achieve significant learning outcomes.[32] An interesting video[33] shows how to create gravitational waves on a membrane using an electric drill and castor wheels to simulate a high speed binary orbit.

*3.1.9 Limitations of this model*
The above discussion illustrates one limitation of the elastic membrane model. The analogy is suitable to show orbital motions, curved space, photon trajectories etc. However discussion of its serious limitations is an essential part of the program. Despite being designed to illustrate the conception of gravity put forth by Einstein, all the gravitational effects in the membrane model are created by Newtonian gravity. The model creates two-dimensional spatial curvature when gravitation as we experience it on Earth is entirely created by the warping of time. Spatial curvature while visible in starlight deflection and geodetic precession has almost no effect for gravity on Earth. The elastic membrane uses spatial curvature to mimic the temporal curvature (usually called warped time), but then we also use it to discuss spatial curvature effects such as precession. To introduce the origin of gravity in Einsteinian physics, it is necessary to focus on gravitational time dilation and warping of time. We have addressed this issue in a separate paper.[34]

## 3.2 Geometry in curved space: non-Euclidean geometry
We now consider a different model that we use to deepen students understanding of curved space and the associated non-Euclidean geometry. The history of curved space begins with Carl Friederich Gauss (1777 – 1855). According to his "Theorema Egregium", the curvature of any surface can be determined by measuring angles and distances within that surface. He invented an instrument called "heliotrope" which used sunlight to construct light beam triangles between three mountain peaks.[35] The purpose was to see whether the sum of angles equalled 180 degrees.[36] Now that we know the tiny magnitude of local space-time curvature, it is not surprising that Gauss failed to measure any difference from 180 degrees. To illustrate non-Euclidean geometry, we use woks and balloons.

*3.2.1 The meaning of a straight line*
The concept of a straight line, which people take for granted, is actually a rather sophisticated concept. This discussion is designed to introduce students to the concept of a straight line. First, we ask students to consider optical sighting as a method of determining whether a line is straight. Since nothing moves straighter than light, it would seem reasonable to use light as a measure of straightness. However, straight lines are defined as the shortest path between two points. This exercise is intended to allow students to experience these two ways of defining straightness in two-dimensional curved space.

Our model for curved space is the surface of an upturned metallic wok. To construct a triangle, students need to draw three straight lines which can be created in two ways a) a tightly stretched string is used to define the shortest distance. b) a "fence line" is created by sighting from two magnetic poles to the next one, thereby defining straightness by local extrapolation of light paths. This is the method used by land surveyors. Both methods are as shown in Figure 7(a). Care must be taken to ensure that the string takes the shortest path, because it can deviate due to friction.

*3.2.2 Geometry on woks*
Prior to the investigating geometry in curved space, we ask students to draw triangles of various sizes on a flat paper using a ruler, and to measure the sum of the angles of each triangle using a protractor.

They will find that the sum of the angles of a triangle is always 180 degrees. After doing this, we ask students to construct triangles of various sizes on upturned woks using strings and magnetic poles (as shown in Figure 7(b)), and to measure the sum of the angles using an adjustable 360 degrees protractor. For small triangles, students will find that the sum of the angles will be approximately 180 degrees, but will exceed 180 degrees for larger triangles.

Students can measure these values using a piece of string and a protractor. Students can plot this relationship between the sum of the angles of a triangle and its perimeter (see figure 7(c)).

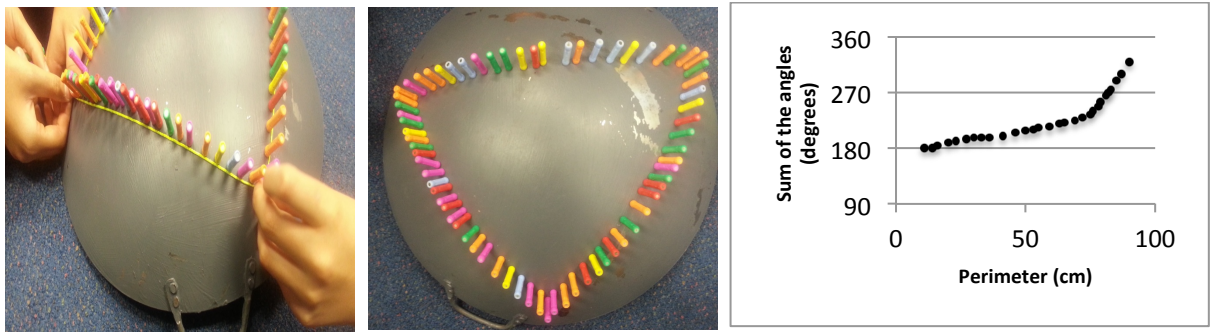

Figure 7(a). Students comparing the two methods of defining a straight line in curved space. Magnets are used to create sightlines. The stretched string confirms that this is the shortest path.

Figure 7(b). A completed triangle. Students measure the angles with protractor, and the perimeter with a piece of string.

Figure 7(c). The data shown here has been gathered from traingles of increasing perimeters. Starting from 180 degrees for a triangle of perimeter 10cm, the sum of the angles of a triangle in curved space increases with size (perimeter). Note that curves like this depend on the shape of the triangle.

The sum of the angles of a triangle can be used to measure the shape of curved space (since the curvature of space determines how the sum of the angles and the perimeter of a triangle are related). This idea can be extended to observations of gravitational lensing for light beams passing by the Sun. This deflection of starlight past the Sun (1.75 arc seconds) corresponds to a sum of angles of a triangle large enough to encompass the Sun equal to 180 degrees plus about three times the deflection angle, about five extra arc seconds!.

### 3.3 Special Relativity

Einstein's theory of special relativity[37] is encompassed within general relativity. However, it is useful to address its key concepts – spacetime and the universality of the speed of light – separately. Special relativity predicts the well-known fact that no object can exceed the speed of light. Relativity describes how moving clocks run slower and moving lengths contract according to time dilation and length contraction respectively. Moving objects also get heavier through relativistic mass increase.

We have chosen to introduce special relativity by focusing on the universality of the speed of light and how it is enforced. We do this by generalising the idea of terminal velocity. Relativity shows that the universe enforces this speed limit by increasing the mass, and therefore the amount of energy required to accelerate a moving object. As an object reaches relativistic velocities, more and more energy is used to increase its mass, and thus it becomes more resistant to acceleration. As the object approaches the speed of light, its mass approaches infinity, barring it from ever reaching this universal constant. This relativistic mass increase is permitted by the interchangeability of energy and mass, a concept encapsulated by Einstein's famous equation $E = mc^2$. The phenomenon is observed in particle accelerators such as those at CERN. In the text below we expand on the above argument.

First we ease students into this counter-intuitive concept of a universal speed limit with an experiment involving falling objects that reach terminal velocity after a small drop. Using balloons of different mass, students can appreciate how a falling object accelerates until it reaches terminal velocity (see Section 3.3.1). Students are then asked to consider physical constraints that limit the object to its terminal velocity. We then apply this understanding to other more complicated examples of objects reaching terminal velocity, for example, a car accelerating down a highway (with no speed limit). Every system of moving objects has a terminal velocity determined by a variety of factors. For a racing car, it may be a balance of fuel availability and wind resistance. For a boat, it may be energy lost in the waves it creates. It is very useful for students to discuss these factors.

A falling object (mass m) in the atmosphere expends energy mg for every meter fallen. For example, 1 kg object that falls 1000 meters has expended 9.8 KJ of energy. Only a small proportion of this has turned into kinetic energy. The rest is lost to friction, determined by surface area of the object and the viscosity of air. Terminal velocity has been enforced by friction. This is illustrated in Fig 8(a)

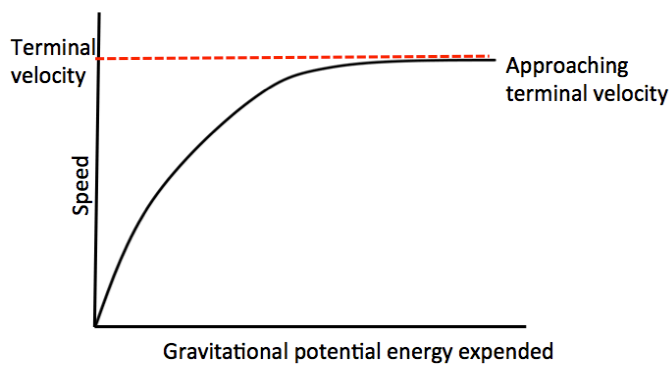

Figure 8(a)

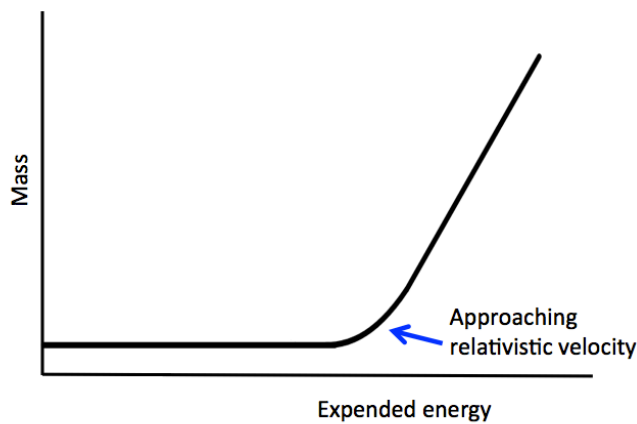

Figure 8(b)

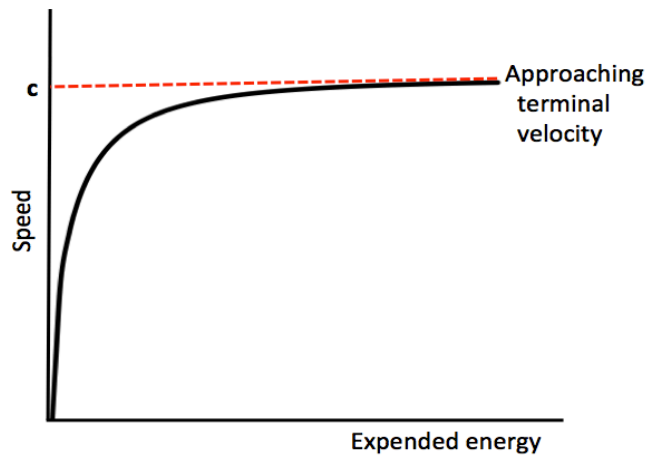

Figure 8(c)

Figure 8. The connection between classical terminal velocity and the speed limit of the universe. Figure 8(a) refers to a falling object in the atmosphere, where the energy source is gravitational potential energy. At low speed, air resistance is negligible and $v$ increases as $\sqrt{E}$ from the formula for kinetic energy $E = \frac{1}{2}mv^2$. At higher speed, acceleration reduces due to viscosity and eventually the object reaches terminal velocity. In Figure 8(b) we consider an object accelerated by an external source of energy such as that proposed by project Starshot.[38] Using the standard relativistic mass formula, the total mass becomes $m_0 + \frac{E}{c^2}$. The mass increases with energy as shown. In Figure 8(c) the speed initially increases proportional to $\sqrt{E}$ as shown in Figure 8(a), but now reaches a terminal velocity c due to the relativistic mass increase. Figure 8(c) was calculated using the standard relativistic gamma factor[39] which we do not present here.

For an object accelerated to relativistic speeds, applied energy begins to increase the mass of the object as well as its speed as illustrated in Figure 8(b). As the body reaches the speed of light, an increasing fraction of the applied energy is used to increase the mass of the body at the expense of increasing speed. Close to the speed of light, almost all the applied energy is used to increase the mass. The body can no longer accelerate. This is illustrated in Fig 8(c). The speed limit of the universe has been enforced by the relativistic mass increase of the moving object.

Having understood the speed limit of the universe, we ask students to consider how to define spacetime when space and time are measured by different and arbitrary human derived units such as seconds and meters. First we introduce the familiar concept of measuring distance with time, as illustrated by advertising signs such as "McDonalds 5-minutes ahead". Students recognize that such statements imply knowledge of a standard velocity. We have found that having grasped the universality of the speed of light students quickly realise this is the universal speed to connect space and time. Thus time can be measured in meters by multiplying time intervals by the speed of light. Equally distances can be measured in time units such as light seconds or light years. This facilitates the higher level understanding needed to understand the origin of gravity.[40] This topic is beyond the scope of this paper.

*3.3.1 Measuring the terminal velocity of a balloon*
The measurement of terminal velocities can be easily accomplished by dropping weighted balloons and measuring their fall using modern smartphones capable of displaying video files frame-by-frame. We mark 10 cm vertical height divisions on a wall so they can be easily seen from a few meters away. One student releases a balloon vertically from a reasonable height (sufficient to allow acceleration to terminal velocity) while the other student records its fall. The balloon must fall vertically as any sideways drift corrupts results. We ask students to video the fall of balloons containing varying amounts of water (~10g - 200g) to obtain varying durations of acceleration and terminal velocity values.

From the video files students create space-time graphs (distance fallen versus time). Treating the initial drop point as the origin (i.e. height equal to zero), and plotting the height (as the distance of the balloon from the origin) of the balloon at each frame, a distance versus time graph such as Figure 9 can be easily created. The frame rate of their cameras (usually 25, 30 or 120 frames per second) must be known to convert 'frame number' to 'time'.

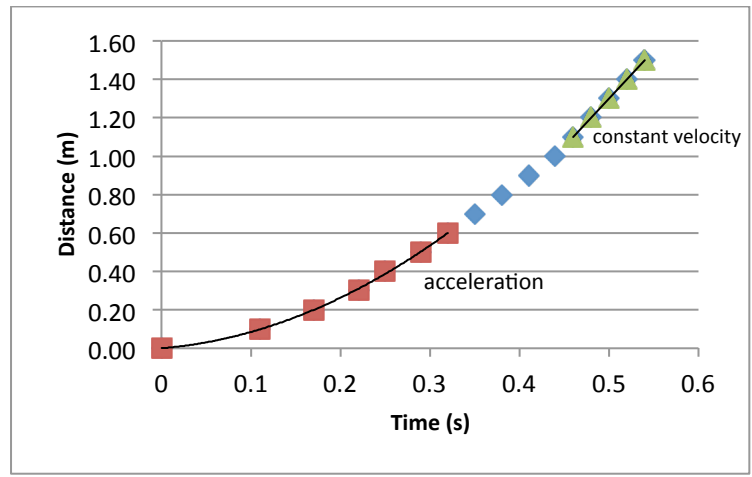

Figure 9. Graph of distance versus time for a balloon dropped from a stationary height (1.5 m). The balloon accelerates until reaching terminal velocity at 0.5 seconds, at which point it falls with a constant velocity.

Students can then use this graph to produce a plot of velocity versus time, which will reveal information about the terminal velocity of the balloon. Students can see that the balloon quickly reaches a maximum velocity that is independent of the initial drop height. Having completed the above activity students are equipped to grasp the speed limit of the universe ideas discussed above.

## 4. Conclusion

The models and analogies described in this paper have been developed and refined in conjunction with many classroom programs over the past five years. We have found them to be effective explanatory tools for teaching concepts of relativity in the high school classroom. In addition to this, the interactive and engaging nature of these activities were found to enhance students' enjoyment and attitudes toward the sciences, leading to more positive learning outcomes.

More research is needed to evaluate the efficacy of these activities in leading to long-term positive learning outcomes for students. Furthermore, future research must investigate and develop ways in which these activities can be incorporated into a formal term-based school setting, complete with the corresponding learning material, suitable testing measures, and into the wider curriculum as a whole. Part 2 of this series of papers considers models and analogies for teaching quantum physics and Part 3 reviews some of the research findings from our programs.

## 5. Acknowledgements

This research was supported by the Australian Research Council, the Gravity Discovery Centre and the Graham Polly Farmer Foundation. We wish to thank Ron Burman, Elaine Horne, David Treagust, Grady Venville, Yohanes and Si Yu for help and encouragement in this work.

**Appendix 1**

Here we outline the materials required to construct the various models and activities described in this paper:

**1) Space-time simulator:** There are many ways to construct this apparatus, however the one described in this paper was constructed using a lycra membrane with dimensions 1.2m x 1.2m, a square wooden frame with dimensions 3m x 3m (or 20 tent poles joined in a large ring), 100 golf balls, steel ball bearings or various size, and spring loaded toy cars with fixed steering.

**2) Geometry on woks:** We used steel woks of different diameters (30 - 35 cm) and curvatures, toy magnets, protractors (ideally a folding ruler with a 180 degree protractor attached to the joint), strings, and graph paper.

**Appendix 2**

Kepler's third law of elliptical orbits can be derived directly from Newton's law of gravitation by equating with and the centripetal force due to circular motion.

$$F = \frac{Gm_1m_2}{r^2} = \frac{m_1 v^2}{r} \qquad (A1)$$

Substituting the velocity of circular motion: v = 2πr/T, we obtain

$$\frac{Gm_1m_2}{r^2} = \frac{4\pi^2 r^2 m_1}{T^2 r} \qquad (A2)$$

Simplifying, and arranging for period in terms of radius, we get

$$T^2 = kr^3, \quad where \; k = \frac{4\pi^2}{Gm_2}$$

This is the standard form of Keplers third law $T^2 \propto r^3$.

**Appendix 3**

The method outlined in Appendix 2 can be applied in the same way to derive a modified Kepler's third law for orbits (period in terms of radius) on our space-time simulator.

As described in section 3.1.1, the force law for 'gravitation' on the membrane is given by:

$$F = \frac{Am_1m_2}{r^{0.6}} \qquad (A3)$$

Following the procedure in Appendix 2 by equating this force with (A2), we obtain

$$\frac{Am_1m_2}{r^{0.6}} = \frac{4\pi^2 r^2 m_1}{T^2 r}$$

Simplifying and arranging for period in terms of radius, we obtain

$$T^2 = \frac{4\pi^2}{Am_2} r^{1.6}$$

Thus orbits on a membrane surface satisfy a modified form of Keplar's third law given by $T^2 \propto r^{1.6}$

Hence the period-radius curve in Figure 5 has the form $T \propto r^{0.8}$.